\documentclass[%
 reprint,
nofootinbib,
 amsmath,amssymb,
 aps,prl
]{revtex4-2}
\usepackage{graphicx}
\usepackage{dcolumn}
\usepackage{bm}
\usepackage{lineno}
\usepackage{hyperref}
\usepackage{latexsym,amsmath,amssymb}
\usepackage{bbm}
\usepackage{physics}
\usepackage{color}
\usepackage{tikz}
\usepackage{subcaption}
\usepackage{tikz-feynman}
\usetikzlibrary{arrows}
\usepackage{graphicx}
\usepackage{multirow}
\usepackage{cancel}
\usepackage{diagbox}
\usepackage{makecell}
\usepackage{cleveref}
\usepackage{pdfpages} 
\usepackage{pgffor}

\newcommand{\be}{\begin{equation}}
	\newcommand{\ee}{\end{equation}} 
\newcommand{\bea}{\begin{eqnarray}}
	\newcommand{\eea}{\end{eqnarray}}
\newcommand{\bfx}{{\bf x}}
\newcommand{\bfr}{{\bf r}}
\newcommand{\br}{{\bm r}}

\newcommand{\bk}{{\bm k}}

\newcommand{\bq}{{\bm q}}
\newcommand{\bp}{{\bm p}}
\newcommand{\bi}{{\mathrm{i}}}
\newcommand{\me}{{\mathrm{e}}}

\newcommand{\sgn}{{\mathrm{sgn}}}
\newcommand{\nn}{\nonumber \\}

\makeatletter
\AtBeginDocument{\let\LS@rot\@undefined}
\makeatother
\def\supplementfilename{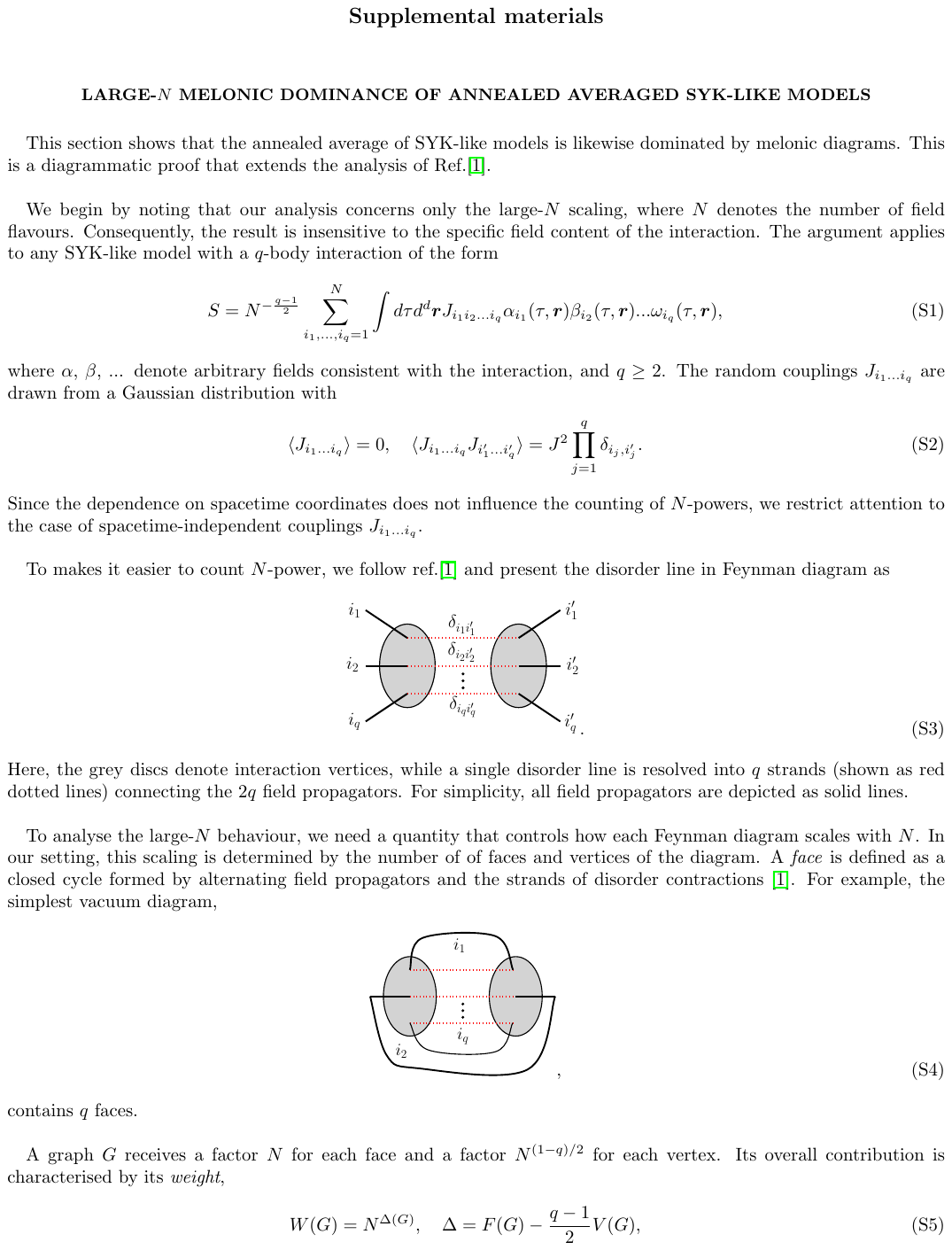} 
\pdfximage{\supplementfilename}
\def\numbersupplementpages{\the\pdflastximagepages}

\newif\ifarXiv
\arXivtrue

\begin{document}

\preprint{APS/123-QED}

\title{ER=EPR and  Strange Metals from  Quantum Entanglement: \\
	  Disorder theory vs quantum gravity}

\author{Sang-Jin Sin}
\email{sjsin@hanyang.ac.kr}
\author{Yi-Li Wang}
\email{wangyili@hanyang.ac.kr}
 \affiliation{Department of Physics, Hanyang University,\\
 	222 Wangsimni-ro, Seoul, 04763, Public of Korea}

\date{\today}

\begin{abstract}
We   give an understanding how  strange metals arise from the  spatially random Yukawa-SYK model based on the wormhole picture and find a parallelism between the disorder theory and quantum gravity. We  start from the observation   that the Gaussian average over the  spatial random  coupling  gives a wormhole,  defined as a mechanism for  long range interaction without causal suppression outside the lightcone. 
We find that    the   large-$N$  limit equivalence of  the quenched and annealed  averages   provides a field theory version of the ER=EPR.  
Since the wormhole establises  momentum exchanges over arbitrary distance
without causal suppression, it    provides a mechanism of the planckian dissipation. It also tells us why  SYK-like  models describe   strongly interacting systems  even in the  small coupling case. 
We classify the disorder samples into two classes: I) 
 spatially random coupling with wormholes and no information loss, 
 II) spatially uniform coupling with  decoherence. 

 %
 
 \end{abstract} 

\keywords{ Yukawa-SYK model, mean field theory, disorder, wormhole, entanglement, ER=EPR, } 
\maketitle

\paragraph{\bf Introduction}
The strange metal, characterised by a linear-$T$ resistivity, has been one of  the most profound puzzles in modern physics \cite{Lee2006,Anderson2017,Lee2018,Greene2020,Varma2020,Hartnoll2022,Phillips2022} due to the lack of theory, despite its universal appearance in strongly correlated metalic system. 
 Recently, it was shown that a  class of  simple  models  \cite{Patel2023,Wang:2024utm,Wang:2025oiz} can  produce the linear-$T$ resistivity in $(2+1)$-dimension based on the disorder field theory.  
These models resemble the Sachedev-Ye-Kitaev (SYK) model \cite{Sachdev1993,Kitaev2015,Chowdhury2022} in
the sense  that    interaction is  all to all and   randomised   over the  space dependent coupling.   
The simplest  such  model  is the Yukawa-SYK model whose  action consists of    Yukawa interaction:
\begin{eqnarray}\label{eqn:dis}
	{\cal L}_{\text{int}}=g_{ijk}(\br)\psi^{\dagger}_i\psi_{j}\phi_k/N\equiv g_{ijk}(\br)H_{ijk}(\br,t),
\end{eqnarray}
 with  spatially random coupling satisfying 
   \bea    \langle g_{ijk}(\br)\rangle &=&0,  \label{eqn:average}\\ 
	\langle g^*_{ijk}(\br)g_{i'j'k'}(\br')\rangle &=& g^2\delta(\br-\br')\delta_{ii'jj'kk'}. \label{eqn:fluctuation}
\eea 
 The theory is defined by a quenched disorder with Gau\ss ian distribution.
 Each field is labeled with a color index $i=1,...,N$, ensuring that the vertex correction of  the theory is well controlled in IR limit \cite{Esterlis2021}.
The  universality of this model was  examined  by replacing the Yukawa interaction by a vector interaction:  
$\phi\psi\psi\to A^{\mu}_{ext}\psi\partial_\mu\psi$ \cite{Wang:2024utm}, which turns out to be the only alternative. It was also pointed out that the inverse Hall angle does not have $T^2$ behavior. \\

The origin of strange metals  is widely believed to lie in many-body quantum  entanglement \cite{Chowdhury2022}. However, it  is  unclear how  
   SYK-rised models formulated  in terms of the   disorder  can generate quantum critical point which is   rooted in quantum coherence \cite{Patel2023,Wang:2024utm, Wang:2025oiz}.  Entanglement alone does not account for the emergence of the strange-metal phase: while many interactions can generate entanglement, few lead to this behaviour. What, then, is the distinctive feature of the Yukawa–SYK model that produces it? Addressing this question is the central aim of this work.
 
The key observation of this work is that the spatially random correlation condition \eqref{eqn:fluctuation} effectively acts as a wormhole, enabling long-range momentum transfer and generating quantum entanglement. Thus, the role of spatially random couplings is not to induce decoherence, but rather to create entanglement that coherently links different regions of the sample into a single quantum state. This mechanism parallels phenomena in quantum gravity, where wormholes serve to connect disconnected regions of spacetime into a unified geometry \cite{VanRaamsdonk:2010pw}.
    \vskip .3cm 
\paragraph{\bf Yukawa-SYK model} 
We start with an action  with  disorder,  $S_{\text{tot}}=S_\psi +S_\phi +S_{\text{int}}$, where
\bea \label{eqn:action}
	\small 
	S_\psi &=&
	 \int d\tau d^2\br \Big[\sum_{a=1}^N \psi_a^{\dagger}(\br,\tau)\left(\partial_\tau-\frac{\nabla^2}{2m}-\mu\right)\psi_a(\br,\tau)\nn
	&&+\sum_{a,b=1}^N V_{ab}(\br)\psi^{\dagger}_a(\br)\psi_b(\br)\Big], \\
 	S_\phi&=&\frac{1}{2}\int d\tau d^2\br \sum_{a=1}^N \phi_a \left(-\partial_\tau^2-\nabla^2+m_b^2\right)\phi_a(\br,\tau), 
\eea 
and  	$S_{int}=\int d^2\bfr d\tau {\cal L}_{int}$ with   
${\cal L}_{\text{int}}$ given  by  \eqref{eqn:dis} together with  condition  \eqref{eqn:fluctuation}.
The potential $V_{ab}(\br)\equiv\sum_{i}V_{\text{imp}(ab)}(\br-\br_i)$ is due to  imperfections at $\br_i$, which satisfies $	\langle V_{ab}(\br)\rangle=0$ and 
\begin{eqnarray}
 \langle V^*_{ab}(\br) V_{a'b'}(\br')\rangle=v^2\delta(\br-\br')\delta_{aa',bb'},  \hbox{ with } v\in\mathbb{R}. 
\end{eqnarray} 
The path integral quantization of this system is given by $\mathcal{Z}\equiv\int D[\Psi]\exp(-S_{\text{tot}})$, with $ D[\Psi]=D[\psi,\psi^\dagger] D[\phi] $. 
The   free energy in the  \emph{quenched} average  is defined by   
\begin{eqnarray}\label{eqn:Zq}
 \langle\ln\mathcal{Z}\rangle_{\text{dis}} =\int D[g]P[g] \left[ \ln(\int D[\Psi]  e^{-S_{\text{tot}}[g,\psi,\phi]})\right],  \; 
\end{eqnarray}
while  that in the \emph{annealed} average is computed by 
\begin{eqnarray}\label{eqn:Za}
 \langle\mathcal{Z}\rangle_{\text{dis}}  = \int D[\Psi] \int D[g]P[g]  e^{-S_{\text{tot}}[g,\psi,\phi]}.  
\end{eqnarray}
In both cases, $P[g]$ represents the Gau\ss ian distribution $P[g]=\exp{-\int  \sum_{ijk}g^2_{ijk}(\bfr)/2g^2}$. 

\begin{figure}[htbp]  
	\centering
	\includegraphics[width=0.3\textwidth]{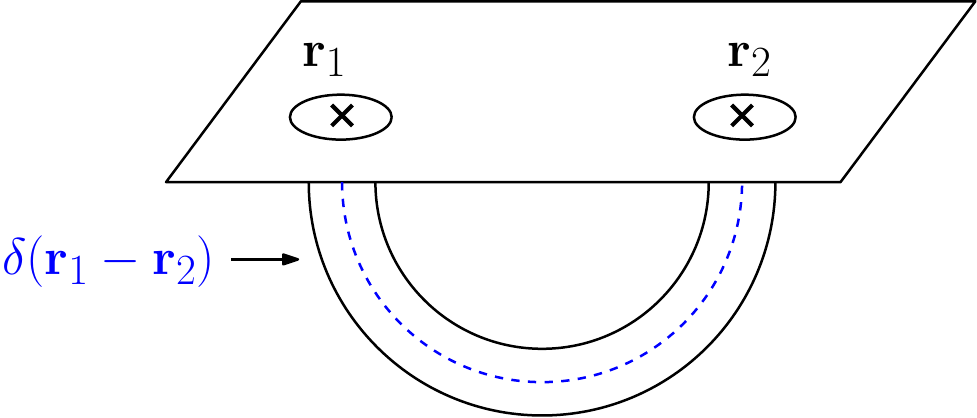}	
	\caption{\small  Two distant spatial points $ \br_1$ and $ \br_2 $ are connected by a wormhole whose throat is of zero length.
		\label{fig:wormhole}}
\end{figure}

\paragraph{ \bf Wormhole in Yukawa-SYK model:}
The free energy in the quenched average $F_q=	-\langle \log Z\rangle_{dis} $    contains only connected diagram  
\begin{eqnarray}\label{eqn:Z}
	\small 
	F_q=-\sum_{n=0}^\infty\int_{\bfx_1 ... \bfx_n} 
	\Big	\langle 
	\langle \prod_{a=1}^{n}  g_{ijk}(\bfx_{a})H_{ijk}(\bfx_a) \rangle_{\Phi,c} \Big\rangle_{dis} ,
\end{eqnarray}
where $\bfx=(\br,t)$ and 
$	 \langle ... \rangle_{\Phi,c}$ is connected diagrams for fixed coupling, while 
$	 \langle ... \rangle_{dis}=\int Dg  ... e^{-S_g} $ is the disorder average. 
Let's  consider the non-trivial lowest order diagram coming from a  $n=2$ term,   
$$\int d\bfx_1 d\bfx_2  \langle  g_{ijk}(\br_1) g_{i'j'k'}(\br_2)\rangle_g  \left\langle  H_{ijk} (\bfx_1)H_{i'j'k'}(\bfx_2) \right\rangle_\Phi. $$ 
Notice that in the perturbation expansion, the correlation functions of $g(\bfr)$ and $\Phi(\bfx)$ are completely  independent of each other.  
The $n=2$  term is graphically represented  as the first diagram in \eqref{eqn:2pt}, where the blue dashed line represents  $\langle  g_{ijk}(\br_1) g_{ijk}(\br_2)\rangle_{dis} $. 
\begin{eqnarray}\label{eqn:2pt}
	\includegraphics[width=.2\textwidth]{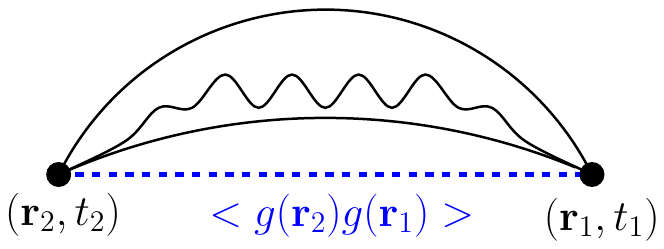}, \qquad 
	\includegraphics[width=.15\textwidth]{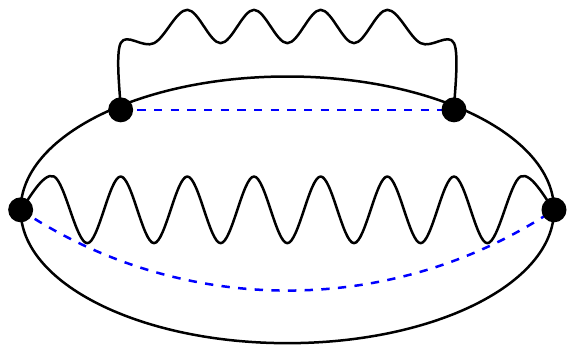}
\end{eqnarray}
According to eqn.\eqref{eqn:fluctuation}, the spatial positions of two vertices connected by the blue dotted line are effectively identified. In the absence of such a connection, propagators of dynamical fields are suppressed outside the light cone by a distance-dependent factor,
 \be e^{-mr}/r^\alpha,\ee
  which we call  as causal suppression.  
  {Notice that bosons in the critical case have relativistic dispersion and $m\to 0$.} 
In contrast, the dotted line eliminates this suppression entirely, regardless of the separation between the vertices. In a theory without spatially random disorder, such behaviour could only occur through a geometric shortcut, akin to a wormhole, connecting the two points (see Fig.~\ref{fig:wormhole}). We therefore refer to this dotted line as a field-theoretic wormhole, or simply a wormhole when no ambiguity arises. A more involved example appears for $n = 4$ in the second diagram of eqn.\eqref{eqn:2pt}, which contains two such wormholes.

\paragraph{    Wormhole and the planckian dissipation:}
Introducing spatial random disorder significantly alters both propagators and self-energies, leading to the remarkable change in resistivity from $\rho=\rho_0+T^\alpha$ with $\alpha=2$ or $4/3$ to the linear form $\alpha=1$, as summarized in Ref. \cite{Wang:2024utm}. This phenomenon can be understood through the presence of the wormhole: it enables particles separated by arbitrary distances to interact with non-decaying amplitude, inducing  both long-range entanglement and momentum transfer. We therefore propose that this ``field-theory wormhole'' underlies Planckian dissipation and is the very  mechanism responsible for strange-metallicity.\\

\paragraph{  Disorder vs entanglement.}
Our wormhole arises as a consequence of the spatially random coupling $g_{ijk}(\bfr)$. A disordered system is defined through an ensemble average, which effectively introduces interactions between different samples. From the perspective of any single sample, this connects the system to an environment, rendering it an open system. As a result, one might expect disorder to induce decoherence. This raises an important question: if strange-metal behaviour is disorder-driven, how can it still be attributed to quantum criticality?
This issue parallels Hawking’s argument in quantum gravity \cite{Hawking1987}, where wormholes were suggested to cause information loss by providing a channel for information to leak into another universe. We address this concern by noting that, after integrating out the random disorder field, possible in the annealed average, the source of decoherence is converted into a non-locality in the effective quantum field theory.
\vskip .3cm
 
 \paragraph{\bf Equivalence of the quenched and annealed averages}
The idea is to go to  annealed average   using the equivalence of two  averages in large-$N$ limit.   
Although  there is a subtlety associated with the boson degree of freedom \cite{Baldwin2020} for  the equivalence of the quenched  and annealed averages, there are   reasons    to trust the equivalence \cite{Shi2023,Benini:2024cpf} in our Yukawa-SYK model. We can actually  demonstrate it   at the level of the free energy and    conductivity,  the most important cases for us. 

 \paragraph{Equivalence of the free energy: }  
 {The annealed average, $\langle Z[g]\rangle$, yields an effective interaction
 \begin{eqnarray}\label{eqn:int}\small 
 	\!\!\!\! S'_{\text{int}}\!\!=-\frac{g^2}{2N^2}\!\!\int \!\! d\tau_1 d\tau_2 d^2\br \!\!\!\!\sum_{a,b,c=1}^N\!\!\!\! H^\dagger_{abc}(\bfr,\!\tau_1) H_{abc}(\bfr,\!\tau_2) , 
 \end{eqnarray}
 { with } $H_{abc}(\bfr,\tau_1)$ given in \eqref{eqn:dis}.
 The corresponding Feynman rule identifies the spatial positions of the two vertices, represented by a dashed-dotted line:
 \begin{eqnarray}\label{eqn:fr}
 	\includegraphics[width=.3\textwidth]{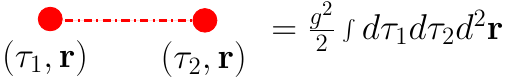}.
 \end{eqnarray}
 This effective vertex is bi-local in time, and the dashed-dotted line plays the same role as the disorder line in the quenched average. Thus, $\langle Z[g]\rangle$ sums over all connected diagrams where these dashed-dotted lines act as $g$-propagators. By contrast, in quenched average, $\langle \ln Z[g]\rangle$ excludes diagrams that become disconnected once the disorder lines are removed. For instance, the diagram 
 \begin{eqnarray}\label{eqn:pseudo}
 	\includegraphics[width=.1\textwidth]{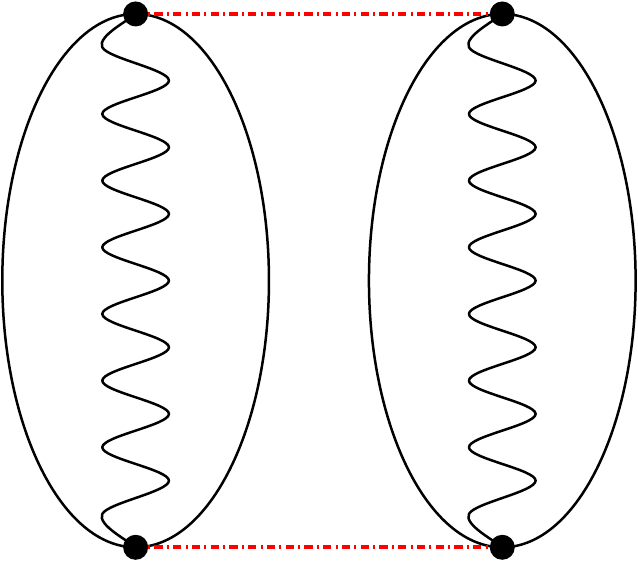}
 \end{eqnarray}
 exists only in the annealed average \cite{Shi2023}. We refer to such diagrams as `pseudo-connected', since their connectivity depends entirely on disorder-line connections rather than matter-field propagators.
 As $N \to \infty$, the quenched average is dominated by melonic diagrams, as in standard SYK models. These diagrams correspond to the replica-diagonal saddle in the replica trick \cite{Shi2023}. Extra pseudo-connected diagrams in the annealed average are always subdominant because they involve fewer independent flavour sums, due to the way disorder lines connect vertices. 
 {To see this roughly, note that a disorder line imposes a constraint of the form $\delta_{ii',jj',ll'}$, contracting the flavours of the fields it connects. In melonic graphs, each disorder line typically shares endpoints with at least two field propagators, reducing the scaling by at most a factor of $1/N$. By contrast, in pseudo-connected diagrams, each disorder line links six field propagators that would otherwise carry independent flavours, resulting in a stronger suppression by $1/N^3$.}
 Therefore, at each order of $g^2$, the annealed average admits all Feynman diagrams present in a quenched average, while receiving extra pseudo-connected diagrams, all of which are subdominant at large $N$ \footnote{A diagrammatic proof of melonic dominance in the annealed average is given in the Appendix.}. 
 Consequently, under replica symmetry, the free energy in both the quenched and annealed averages is governed by the same set of melonic diagrams, establishing their large-$N$ equivalence \footnote{It was shown in \cite{Esterlis2021} that the free energy and entropy remain finite under replica symmetry, so symmetry breaking is unnecessary at this stage.}. This result agrees with the analysis on the spatially uniform coupling given in ref.\cite{Shi2023}.
 }


{Although the theory \eqref{eqn:int} is dominated by melonic graphs as $N \to \infty$, it differs fundamentally from the SYK-like Gurau–Witten model \cite{Witten2019} and other melonic-dominant models without disorder, such as the Amit–Roginsky model \cite{Amit:1979ev,Benedetti:2020iku,Nador:2023inw}. The effective interaction in \eqref{eqn:int} originates from disorder and exhibits the all-to-all structure characteristic of SYK models, whereas both the Gurau–Witten and Amit–Roginsky models are genuinely disorder-free. In particular, the Gurau–Witten model is described as ``SYK-like'' solely because of its melonic dominance at large $N$, not because it shares the same interaction structure.}    
 
  \paragraph{Eqivalence in the conductivity.} 
 Based on the effective action \eqref{eqn:int}, we will show that in the large $N$ limit, annealed system \eqref{eqn:int} should give the same conductivity as the quenched average, at the level of two point function of the current operators (using the replica trick as is demonstrated in refs.\cite{Patel2023,Esterlis2021,Guo2022}). The    bi-locality  in time  turns out to be crutial to get linear-in-$T$ resistivity. 
The full propagators     of   electrons and bosons take the form \cite{Altland2023,Patel2023} 
\begin{align} \scriptsize 
	&G(\bi\omega,\bk)=\frac{1}{\bi\omega-\frac{\bk^2}{2m}+\mu-\Sigma(\bi\omega)},\label{eqn:f-propagator}\\
	&D(\bi\Omega,\bq)=\frac{1}{\Omega^2+\bq^2+m_b^2-\Pi(\bi\Omega,\bq)},\label{eqn:b-propagator}
\end{align}
respectively  with corresponding  self-energies $\Sigma$ and $\Pi$. 
In our model, we again obtain extra graphs for self-energy which do not exist in quenched average, and they are subdominant in large-$N$ limit.
For instance, the action \eqref{eqn:int} allows for `semi-tadpole' diagrams 
shown below,
\begin{eqnarray}\label{eqn:tadpole}
	\includegraphics[width=.1\textwidth]{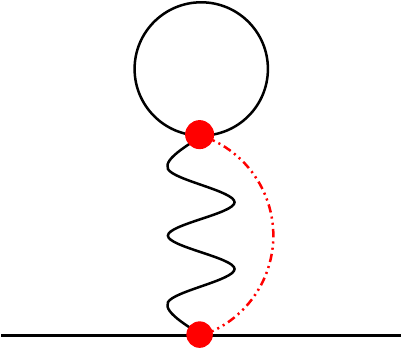} \hskip 1cm 
		\includegraphics[width=.16\textwidth]{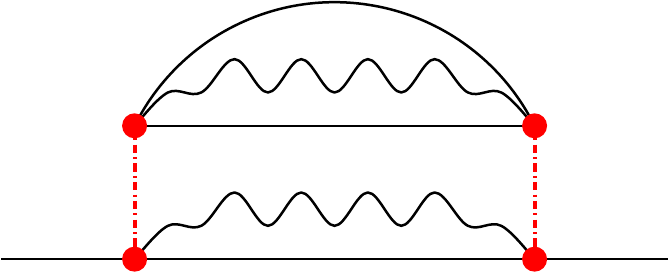}.
\end{eqnarray}
Here the solid lines and wavy line represent fermion propagator and boson propagator respectively.
Left Feynman graph of \eqref{eqn:tadpole} is a tadpole diagram in time, as two sub-vertices are temporally separated, but it is not a tadpole diagram in space, since sub-vertices have the same spatial location. Due to such difference, diagram of this type is termed as semi-tadpole in this article. In QFT, tadpoles are vacuum graphs of no consequence unless  we are   interested in spontaneous symmetry breaking, but  semi-tadpoles can not be so  \textit{a priori}. Thanks to the large-$N$ limit, left graph of \eqref{eqn:tadpole} contributes to electron self-energy  a term of order $\mathcal{O}(1/N)$, while the leading order of self-energy is $\mathcal{O}(1)$. For this reason, all semi-tadpoles can be neglected. \footnote{This diagram can also appear in a quenched average, and it is neglected for the same reason as $N\to \infty$.} Another example from electron self-energies is the right diagram in \eqref{eqn:tadpole}. 
This is a pseudo-connected graph, which is of order $\mathcal{O}(1/N^2)$. Again, all pseudo-connected self-energies are subdominant. Consequently, \emph{only `melonic' graphs contribute to the leading order of self-energies as $N\to \infty$}. The melonic  dominance is a key feature of SYK model \cite{Chowdhury2022}, which is also faithfully captured by SYK-rised theories \cite{Esterlis2021,Guo2022,Patel2023,Wang:2024utm}.

The interaction \eqref{eqn:int}  should therefore exhibit the same properties as a quenched average. We now illustrate this explicitly. 
Firstly, the boson self-energy $\Pi$ can be obtained without knowing the   full propagator $G$ explicitly.
Graphically, $\Pi$ is represented by
\begin{eqnarray}\label{eqn:bsg}
	\includegraphics[width=.4\textwidth]{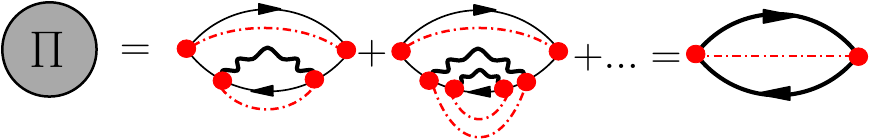}.
\end{eqnarray}
The thin lines represent the bare propagator with $\Sigma=0$ or $\Pi=0$,  and the thick lines are the full propagator \eqref{eqn:f-propagator}. 
Eqn.\eqref{eqn:bsg}   yields
\begin{eqnarray}
	&&\Pi(\bi\Omega,\bq)
	=-2\times\frac{g^2}{2}T\sum_{\omega}\int\frac{d^2\bk}{(2\pi)^2}\frac{d^2\bk'}{(2\pi)^2}\nn
	&&\times\frac{1}{\bi\omega-\frac{\bk^2}{2m}+\mu-\Sigma(\bi\omega)}
	\frac{1}{\bi(\omega+\Omega)-\frac{\bk'^2}{2m}+\mu-\Sigma(\bi\omega+\bi\Omega)}\nn
	&&=-\frac{g^2}{2}\pi\mathcal{N}^2|\Omega|
	\equiv-c_g|\Omega|
\end{eqnarray}
at \emph{zero temperature}, where $\mathcal{N}=m/(2\pi)$ is the Density of State (DoS) at the Fermi energy in $2$D space. We have used the fact that $\sgn(\omega)=-\sgn(\Im{\Sigma(\bi\omega)})$\cite{Esterlis2021}.

\begin{figure}[t!]
	\begin{subfigure}[t]{0.2\textwidth}
		\includegraphics[height=0.8in]{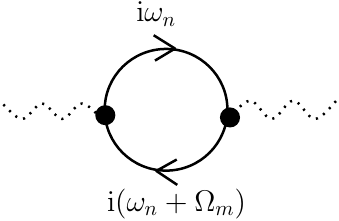} \hskip .5cm 
	\end{subfigure}%
	\begin{subfigure}[t]{0.2\textwidth}
		\includegraphics[height=0.7in]{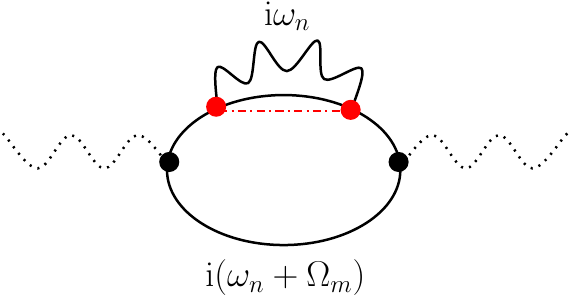}
	\end{subfigure}
	\caption{\small Plarisations contributing to the conductivity. \hskip 1cm 
		(a) The simplest current-current correlator.  \hskip 2cm (b) The polarisation bubble of order $g^2$ 		\label{fig:polarisation}}
\end{figure}

The fermion self-energy $\Sigma$ comprises contributions from the potential disorder $V$ and  from the fermion-boson interaction \eqref{eqn:int}.  
 The impurity scattering can be evaluated after a disorder average via the replica trick, which yields
\begin{eqnarray}
	\Sigma_v(\bi\omega)
	\equiv-\bi\frac{\Gamma}{2}\sgn(\omega),
\end{eqnarray}
with $\Gamma=2\pi v^2\mathcal{N}$ defined as impurity scattering rate \cite{Altland2023}. The other term $\Sigma_g$ at large-$N$ limit reads
\begin{eqnarray}\label{eqn:seg}
	\includegraphics[width=.4\textwidth]{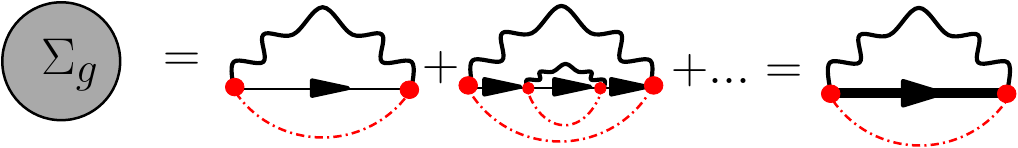}.
\end{eqnarray}
The Fermi surface in this article is designed to be critical,  so we assume $m_b^2-\Pi(0,\mathbf{0})=0$ for the criticality \cite{Esterlis2021,Guo2022}.
One then finds
\begin{eqnarray}
	\Sigma_g(\bi\omega)
	=-\bi\frac{\mathcal{N}g^2}{4\pi}\omega\ln(\frac{\me\Lambda_q^2}{c_g|\omega|}).
\end{eqnarray}
The total electron self-energy,  
$\Sigma=\Sigma_v+\Sigma_g$,  is therefore 
\begin{eqnarray}\label{eqn:selfenergy}
	\Sigma(\bi\omega)=-\bi\frac{\Gamma}{2}\sgn(\omega)-\bi\frac{\mathcal{N}g^2}{4\pi}\omega\ln(\frac{\me\Lambda_q^2}{c_g|\omega|}).
\end{eqnarray}
The self-energy \eqref{eqn:selfenergy} is same as the one found in ref.\cite{Patel2023}, implying that the annealed and quenched averages yield the same results in the large-$N$ limit.
\paragraph{Linear Resistivity.}
Now let us turn on an external electromagnetic field, and compute the conductivity of theory \eqref{eqn:action} from  the \emph{Kubo formula}
\begin{eqnarray}\label{eqn:kubo}
	&&\sigma^{\mu\nu}(\bi\Omega_m)=-\frac{1}{\Omega_m}\left.[\tilde{\Pi}^{\mu\nu}(\bi\Omega)]\right|_{\Omega=0}^{\Omega=\Omega_m},
\end{eqnarray}
as a series of $g^2$.
The current-current correlator, $\tilde{\Pi}_{\mu\nu}$, up to two loops, is  represented by Fig.\ref{fig:polarisation}, where dotted wavy lines stand for external propagators. \\
At the zeroth order, one obtains the simplest polarisation represented by \ref{fig:polarisation}(a), which reads \footnote{A factor $2$ is included due to a summation over spins.}
\begin{eqnarray}
	&&\tilde{\Pi}^{\mu\nu}_0(\bi\Omega_m)
	=-v_F^2\mathcal{N}\delta^{\mu\nu}\frac{\Omega_m}{\Omega_m+\sgn(\Omega_m)\Gamma}.
\end{eqnarray}
Here  again most contributions come from electrons near the Fermi surface.
The next order is illustrated by Fig.\ref{fig:polarisation}(b), which is the contribution from electron self-energy of order $\mathcal{O}(g^2)$. At zero temperature, it yields
\begin{eqnarray}
	\tilde{\Pi}^{\mu\nu}_g(\bi\Omega_m)
	&=&\frac{g^2v_F^2\mathcal{N}^2}{16\pi\Gamma^2}\delta^{\mu\nu}\Omega_m^2\ln(\frac{\me^3\Lambda_q^4}{c_g^2|\Omega_m|^2}).
\end{eqnarray}
Vertex corrections such as Maki-Thompson (MT) graphs and Aslamazov-Larkin (AL) graphs are zero, since the integrands are odd functions of $\bk$ \cite{Patel2023}.
We can thus substitute $\tilde{\Pi}=\tilde{\Pi}_0+2\tilde{\Pi}_g$ to Kubo formula \eqref{eqn:kubo}. Performing an analytical continuation, $\bi\Omega_m\to\Omega$, one obtains \footnote{The extra factor $1/N$ in Eqn.\eqref{eqn:conductivity} is added because the polarisation bubble of the external field is of order $N$.}
\begin{eqnarray}\label{eqn:conductivity}
	&& {\sigma^{\mu\nu}(\Omega)}/{N}\nn
	&=&\delta^{\mu\nu}\Big[v_F^2\mathcal{N}\frac{1}{\Gamma-\bi\Omega}
	+\frac{g^2v_F^2\mathcal{N}^2}{8\pi\Gamma^2}\bi\Omega\ln(-\frac{\me^3\Lambda_q^4}{c_g^2\Omega_m^2})\Big].
\end{eqnarray}
In the absence of a magnetic field, the conductivity $\sigma^{\mu\nu}$ is diagonal, so we drop the superscript $\mu\nu$ from  $ \sigma^{\mu\mu}$.
The resistivity can be obtained using eqn.\eqref{eqn:conductivity}, 
\begin{eqnarray}
	N\rho&=&\Re{\frac{N}{\sigma}}
	\simeq\frac{\Gamma}{v_F^2\mathcal{N}}+\frac{1}{v_F^2\mathcal{N}}\frac{g^2}{8}\Omega+\mathcal{O}(\Omega^2).
\end{eqnarray}
This theory    gives  linear-$T$ resistivity at low temperatures,  same as the quenched average \cite{Patel2023} \footnote{For the general correlation functions,  instead of going into this technical issue further, we simply  assume   the    replica diagonal dominance   in the large-$N$ limit following  \cite{Patel2023,Esterlis2021}, and focus on its consequence.}. Although the calculations of these two average methods are very similar, 
we emphasize again  that  in  annealed average,
  our model \eqref{eqn:int} is a pure  quantum system after  the random coupling is integrated out. 

\vskip .3cm 

\paragraph{ \bf  ER=EPR for spatially random coupling }
The annealed average over partition functions yields an effective action \eqref{eqn:int} containing a six-valent vertex. This bi-local interaction entangles two sets of particles because it imposes only a single momentum-conservation constraint in $S'_{\text{int}}$.
If the total momentum of the Fourier mode of $H_{abc}(\bfr,\tau_1)$ is denoted by $\mathbf{p}$, then that of $H_{abc}(\bfr,\tau_2)$ must be $-\mathbf{p}$.
Thus, particles at $\tau_1$ and those at $\tau_2$ are correlated in momentum space.
The composite system formed by these two sets of particles is in a pure, normalized state $\ket{\Psi}$, which admits the Schmidt decomposition,
$$\ket{\Psi}=\int d\bp\sqrt{P(\bp)}\ket{\bp,-\bp}, \hbox{ with }   \int d\bp P(\bp)=1. $$ 

If we refer to the field-theory wormhole as an Einstein–Rosen bridge prime (ER'), then the equivalence between quenched and annealed averages suggests the relation, $ER'= EPR$, as illustrated below.
 \begin{eqnarray}\label{eqn:qd}
 	\includegraphics[width=.32\textwidth]{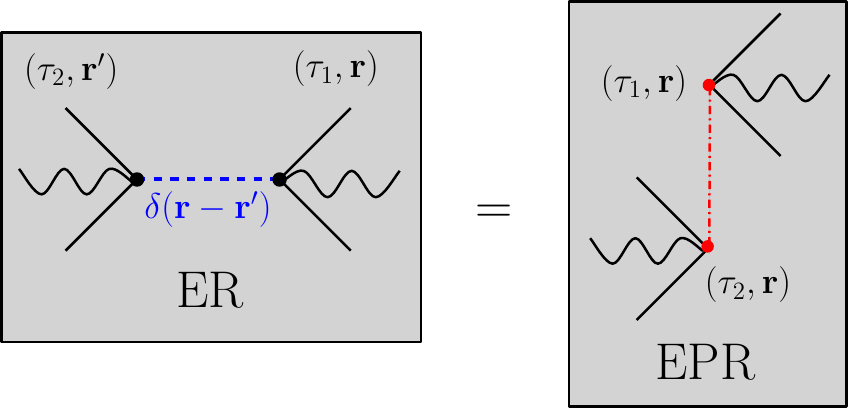}  	 .
 \end{eqnarray}
Namely, each diagram containing a field-theory wormhole in the quenched picture corresponds to a diagram that produces entanglement between the two vertices connected by ER'.
This constitutes a field-theoretic realisation of the conjecture by Maldacena and Susskind \cite{Maldacena2013}, which posits a deep connection between quantum entanglement (EPR) and wormholes (ER).
 

Before proceeding, we emphasise an important subtlety: integrating out the disorder field $g(x)$ does not remove the possibility of  loss of unitarity because the resulting interaction is bi-local in time: notice that  It allows processes where  particles in the original  vertex $H_{ijk}(\tau,\bfr)$,     disappear  at time $\tau_1$ and reappear at $\tau_2$ at the same spatial position. 
	Ealier discussion assumed that they would always reappear before  the measurement. But whatever  late time we may fix,  the measurement time, there is no guarantee that the reappearance time should be ealier than that. 
	 {Such processes generate entanglement between particles at $\tau_1$ and $\tau_2$,  but this only contributes to observable correlations if $\tau_2 < \tau_{measure}$.}
Whether the system remains unitary depends on our ability to localise the entire process within the experimental spacetime region. This, in turn, depends not only on the interaction structure but also on the observables. If both vertices lie within the same universe (i.e., sample), the system remains unitary; otherwise, it effectively behaves as an open system and exhibits decoherence.
{ In other words,  the decoherence issue  reduces to whether the physical system  	allows all the events within our specified experimental zone. 	
	
	For a generic disordered system, the theory involves a sum over a disorder variable, typically a coupling whose value is fixed for each sample. Averaging over disorder amounts to a statistical average across different samples. Due to the Gaussian weight, this averaging effectively introduces correlations between distinct samples. From the viewpoint of a single sample, this can be interpreted as contact with an external environment, leading to decoherence and a loss of unitarity. }

\begin{figure}[htbp]  
	\centering
	\includegraphics[width=0.35\textwidth]{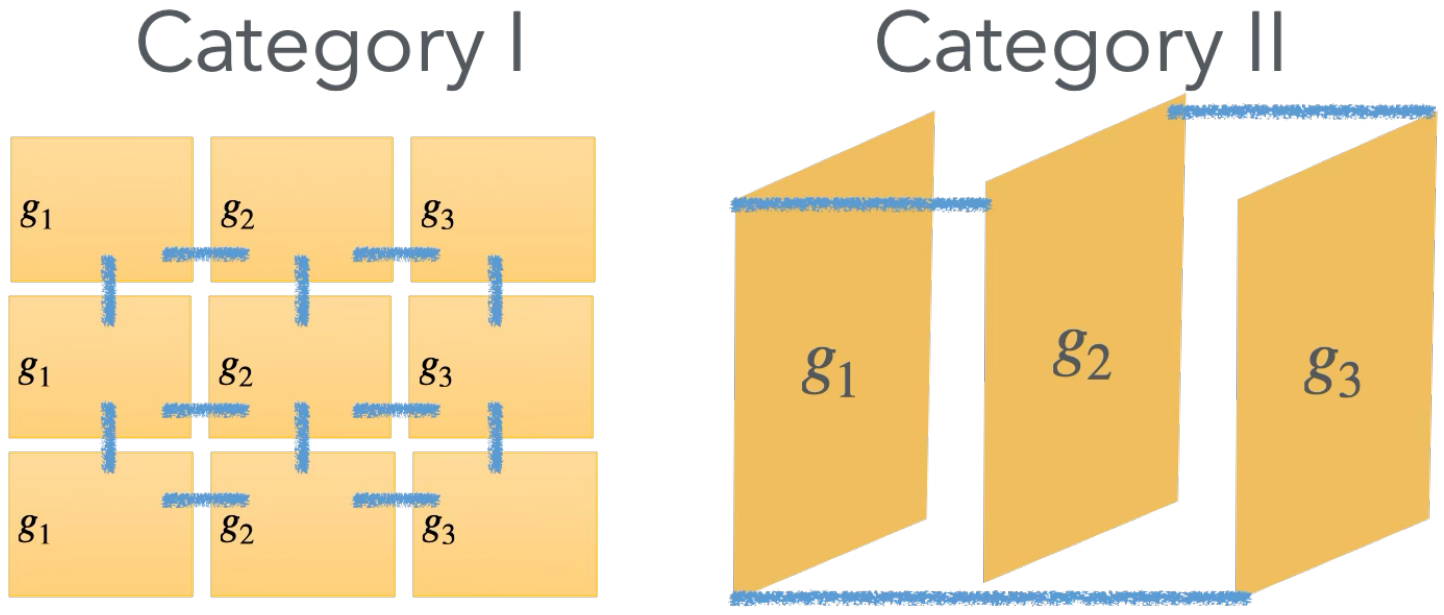}	
	\caption{\small   Left: A sample with spatially random coupling where ER = EPR holds. Right: A sample exhibiting decoherence.
		\label{fig:twocategory}}
\end{figure}

However, when the coupling becomes spatially dependent, the situation changes. This typically occurs in doped systems, where the impurity density—and hence the coupling constant—varies across space. One can visualise this as a single sample composed of many small regions, each with a fixed coupling value.
As pointed out earlier, this leads to a wormhole structure. If a wormhole connects two spatial points $\bfr$ and $\bfr'$, both ends lie within the same sample. That is, all vertices connected by the wormhole remain inside one universe. Under these conditions, there is no information loss: each subsystem is sufficiently localised, so that its recurrence time can be arranged to precede the measurement time. In this setting, the wormhole simply represents entanglement. Here, and only here, the relation ER=EPR holds.
We refer to such systems as category (I), and all others as category (II); see Fig.~\ref{fig:twocategory}.

 {
Category II corresponds to systems with spatially uniform coupling $g$. In this case, the Gaussian average over the sample still establishes a connection between them. The integrating-out process
\begin{eqnarray*}
	&&\int D[g]\exp(-g^2/2+\int dx g\mathcal{O}(x))\nn
	&&\to \exp(\int dx_1dx_2{\cal O}(x_1){\cal O}(x_2)/2)
\end{eqnarray*}
generates an interaction term that connects two vertices, and hence links the samples. Note that the resulting interaction is non-local in both space and time. To accommodate this case within a unified framework, we may introduce a `fuzzy wormhole' whose endpoints are space-filling, such that the two vertices cannot be localised within any specific experimental spacetime region.}
\\

\paragraph{\bf A parallel between disorder theory and quantum gravity}
Hawking once suggested that wormholes could induce decoherence by providing channels through which information escapes \cite{Hawking1987}. This bears a striking resemblance to Category II in our classification. By contrast, the $ER=EPR$ conjecture \cite{Maldacena2013} asserts that every EPR pair \cite{Einstein1935a} is joined by an Einstein–Rosen bridge \cite{Einstein1935}, preserving unitarity—a situation analogous to Category I.
The analogy extends beyond superficial resemblance: in one theory, we integrate over random couplings; in the other, over spacetime metrics.

 {We expect that many phenomena related to entanglement and spacetime structure in quantum gravity have counterparts in Yukawa–SYK models. The correspondence between ER = EPR and the gluing of universes through entanglement are two such examples. Insights gained on one side can illuminate the other: while quantum gravity effects remain largely conjectural, their Yukawa–SYK analogues can be verified through explicit calculation. This perspective may also shed light on other strongly correlated systems, such as superconductors, where time-delay effects play a critical r\^ole in pairing \cite{Tinkham2023}. Persuing these parallels not only deepens our understanding of emergent phenomena but also suggests a unifying language connecting quantum matter and quantum gravity.}\\

\paragraph{\bf Acknowledgments}
The authors would like to thank Mikio Nakahara, Moon-Jip Park, and Jin-Wu Ye for the helpful discussion. This work is supported by NRF of Korea with grant No. NRF-2021R1A2B5B02002603, RS-2023-00218998.


\bibliography{apssamp}
\clearpage
\includepdf[pages=1]{\supplementfilename}
\clearpage
\includepdf[pages=2]{\supplementfilename}
\clearpage
\includepdf[pages=3]{\supplementfilename}
\end{document}